\documentclass[backref=false,pagebackref,bookmarks]{vgtc}                          % final (conference style)
\graphicspath{{figures/}{pictures/}{images/}{./}} % where to search for the images

\usepackage{times}                     % we use Times as the main font
\usepackage{tabu}                      % only used for the table example
\usepackage{booktabs}                  % only used for the table example
\usepackage{lipsum}                    % used to generate placeholder text
\usepackage{mwe}                       % used to generate placeholder figures
\usepackage{listings}
\usepackage{xcolor}
\usepackage{enumitem}

\usepackage{mathptmx}                  % use matching math font

\usepackage[backref=false]{hyperref}

\onlineid{0}

\vgtccategory{Research}

\title{Revisiting put-that-there, context aware window interactions via LLMs}

\author{
Riccardo Bovo\thanks{e-mail: \href{mailto:rb1619@imperial.ac.uk}{rb1619@imperial.ac.uk}} \\
\parbox{1.4in}\scriptsize Imperial College London, UK
\and
Daniele Giunchi \\
\parbox{1.4in}\scriptsize University of Birmingham, UK
\and
Pasquale Cascarano \\
\parbox{1.4in}\scriptsize University of Bologna, IT
\and
Eric J. Gonzalez \\
\scriptsize Google, USA
\and
Mar Gonzalez-Franco \\
\scriptsize Google, USA
}

\teaser{
  \centering
  \includegraphics[width=\linewidth]{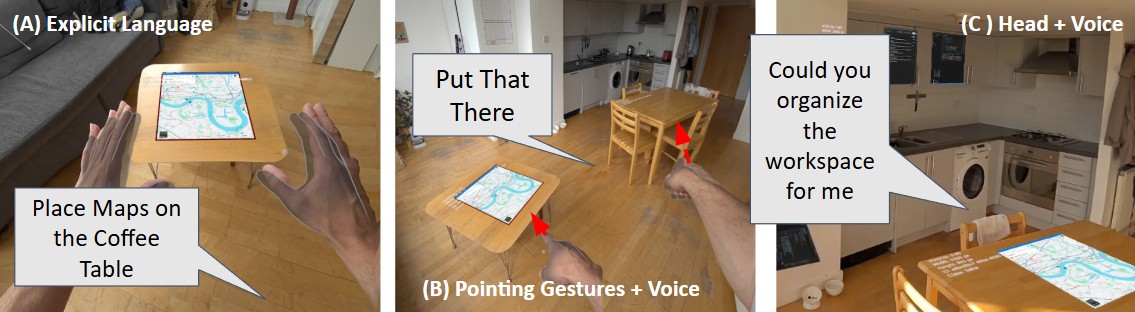}
  \caption{Examples of multimodal interaction styles supported by the system. (A) \textbf{Explicit Language:} The user directly states the action using a complete verbal command. (B) \textbf{Pointing Gestures + Voice:} Deictic terms like "that" and "there" are disambiguated through concurrent pointing gestures. (C) \textbf{Head + Voice:} Implicit commands such as workspace organization are resolved using head orientation and visibility metadata.}
  \label{fig:teaser}
}

\abstract{
We revisit Bolt’s classic \emph{Put-That-There} concept for modern head-mounted displays by pairing Large Language Models (LLMs) with XR sensor and tech stack.  The agent fuses (i) a semantically segmented 3-D environment, (ii) live application metadata, and (iii) users’ verbal, pointing, and head-gaze cues to issue JSON window-placement actions.  As a result, users can manage a panoramic workspace through: (1) explicit commands (\emph{“Place Google Maps on the coffee table”}), (2) deictic speech plus gestures (\emph{“Put that there”}), or (3) high-level goals (\emph{“I need to send a message”}).  %The LLM resolves each request into context-aware window layouts on appropriate, visible surfaces, enabling task-centric interaction without manual window juggling. 
Unlike traditional explicit interfaces, our system supports one-to-many action mappings and goal-centric reasoning, allowing the LLM to dynamically infer relevant applications and layout decisions, including interrelationships across tools. This enables seamless, intent-driven interaction without manual window juggling in immersive XR environments.}


\begin{document}


\firstsection{Introduction}

\maketitle

%% LLMs and XR
The rapid advancement of Large Language Models (LLMs) has revolutionized human-computer interaction, with chat-bots such as ChatGPT \cite{openai2022chatgpt}, Claude \cite{claudeai2023}, and BARD \cite{google2023bard} emerging as the primary interface for engaging with these powerful AI systems. However, as LLMs continue to evolve, their potential extends beyond text-based interactions, particularly in the domain of extended reality (XR) environments. 
%% XR headsets capabilitites
Modern consumer-grade XR headsets,such as the Meta Quest Pro \cite{metaquestpro2022} and Apple Vision Pro \cite{applevisionpro2022}, integrate depth cameras, spatial anchors, and machine-learning pipelines that reconstruct head- and body-movement and classify surrounding elements (walls, furniture, objects) in real time \cite{apple_visionos_movement2025,apple_handtracking2025,meta_movement_sdk_bodytracking2025,metaquestprosceneapi2022,applevisionpro2024}.  
These egocentric sensing capabilities far exceed those of conventional PCs or smartphones, making XR an ideal platform for context-aware multimodal interaction.
%% XR + AI 
XR’s egocentric sensors create a rich context that AI agents can interpret to streamline productivity and attentional tasks. We explore how LLMs levereged in head-mounted displays combine explicit or implicit speech with non-verbal cues and semantic scene representations, revisiting Bolt’s \emph{Put-That-There} paradigm~\cite{10.1145/800250.807503} for today’s XR productivity workflows~\cite{biener2020breaking,pavanatto2021we} and recent AI-driven interaction models~\cite{bovo2025embardiment,bovo2023speech}. 
%% Our contribution
We present a task-centric window-management system that:
\begin{enumerate}[noitemsep,leftmargin=1.2em]
  \item fuses explicit or implicit speech (\emph{“Send a message,” “Put that there”}) with non-verbal cues such as pointing and head-gaze;
  \item selects the relevant application window(s);
  \item grounds user behaviour in the 3-D scene, determining which surface/application the user is referencing (e.g., the coffee table, google maps);
  \item selects a semantically and geometrically appropriate flat surface within the segmented environment for window placement; and
  \item emits action(s) that automatically arrange the window(s) in the user’s panoramic XR workspace.
\end{enumerate}

%By shifting interaction from low-level window repositioning to high-level intent, our approach aims to reduce the overhead associated with manual drag-and-resize operations. 
We discuss the implications of using LLMs, including the transition from traditional one-to-one window actions to one-to-many automated interactions, the shift from explicit placement commands to goal-centric input, and how LLM-based modelling of application interrelationships enhances the adaptability and coherence of XR workspaces. These developments have the potential to significantly reduce cognitive load and improve user efficiency in immersive environments.

\section{Related Work}

\subsection{XR productivity workspaces}
VR head-mounted displays can replace, and vastly expand, conventional multi-monitor setups by rendering arbitrarily large, portable screens at negligible marginal cost~\cite{biener2020breaking,pavanatto2021we}.  This “infinite desktop" unlocks private work-from-home and on-the-go scenarios, mitigating spatial constraints, distractions, and work–life overlap~\cite{fereydooni2020virtual,ofek2020towards}.  Prior HCI studies confirm that mirrored desktop windows~\cite{hoppe2020enabling}, task-aware 3-D screen layouts~\cite{satriadi2020maps,lee2020shared}, and attention-guided cues~\cite{bovo2023speech} can boost knowledge-worker efficiency. However, Pavanatto et al. shows that manually organizing windows in these extensive panoramic workspaces imposes a heavy cognitive overhead~\cite{pavanatto2025spatial}. The authors explore alternative options to reduce the cost of laborious manual window organization in VR spaces. We extend this work by exploring a solution to the same problem in the context of XR, in which 2D window/screen can be anchored anywhere in the environment. We address this gap, arguing that XR’s vast display real-estate and rich egocentric sensing make it the ideal platform for an LLM-driven, task-centric window manager that organizes space on the user’s behalf.

\subsection{XR Multimodal Inputs}
Pointing and gaze have long been recognised as powerful non-verbal cues for resolving referential ambiguity in multimodal dialogue.  Classic work such as \textit{Put-That-There} showed how speech augmented with hand gestures streamlines spatial commands~\cite{10.1145/800250.807503}, and subsequent studies confirmed that users default to pointing whenever verbal description becomes cumbersome~\cite{Wong2010,Bovo2022}.   Beyond explicit gestures, combining speech with continuous attention signals has enabled richer context-aware systems.  Examples include integrating head-gaze with GPS for location-based assistants~\cite{Mayer2020EnhancingWorldGaze}, coupling voice with directive gestures in mobile \emph{VQA} interfaces~\cite{romaniak2020nimble}. Recent XR work further leverages implicit gaze saliency to create or rearrange content covertly~\cite{marwecki2019mise} and to improve head-based attention cues via collaborative speech~\cite{bovo2023speech}. Bovo et al. with \emph{EmBARDiment} dexemplifies how XR systems can tightly couple implicit gaze saliency, verbal input, and contextual memory to create a multimodal interaction loop, where the user’s attention, speech, and visual context are continuously fused to ground AI agent responses in the evolving task environment
~\cite{bovo2025embardiment}. 

Building on this body of evidence, our approach treats pointing and gaze as first-class inputs to a LLM agent.  Rather than using these signals only at the instant of a request, we maintain a lightweight memory of gaze-driven saliency over time.  This enables the LLM to interpret underspecified commands (e.g.\ “put that there”) in light of recent visual context, providing more robust and efficient window-placement decisions than prior one-shot multimodal techniques.

\subsection{XR Content Placement on Physical Environment}
Previous research highlights several advantages of placing XR content on physical surfaces. Consistently anchoring virtual elements to stable, physical surfaces enhances spatial predictability and reduces cognitive load~\cite{davari2024towards}. Furthermore, positioning XR content onto familiar, real-world surfaces facilitates intuitive user interactions, lowering cognitive effort required for interaction~\cite{stanney2021extended}. Ergonomic considerations further emphasize the benefits of anchoring AR content; specifically, attaching virtual elements to stable, flat surfaces such as the floor has been shown to reduce discomfort and motion sickness, thereby supporting safer and more comfortable user interactions~\cite{natarajan2024ergonomic}. Additionally, stable and predictable positioning on flat surfaces reduces eye strain, underscoring their suitability for ergonomically sound virtual content placement~\cite{zhang2024towards}. Building on these insights, our system leverages XR devices semantic and geometric scene understanding to anchor virtual windows onto appropriate flat surfaces. This approach promotes intuitive, ergonomically sound interactions, reducing cognitive load and enhancing user comfort and safety in XR workspaces.

\section{Technical Implementation}

\begin{figure}[t]
    \centering
    \includegraphics[width=\columnwidth]{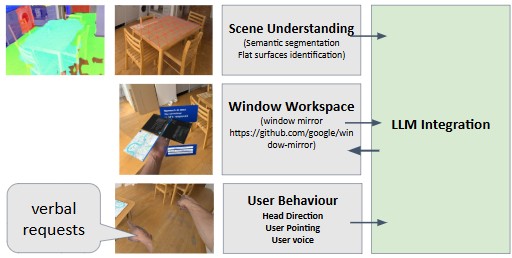}
    \caption{System architecture overview. The LLM receives input from three main modules: (1) \textit{Scene Understanding}, which provides semantic segmentation and identifies flat surfaces; (2) \textit{Window Workspace}, which manages digital window elements using the WindowMirror system; and (3) \textit{User Behaviour}, capturing head direction, pointing, and voice commands. Together, these components allow the LLM to interpret multimodal requests and generate actionable window placement decisions.}
    \label{fig:system-architecture}
\end{figure}

Our system enables users to organize virtual windows within their physical environment through natural multimodal interaction. The core of the system provide the LLM with three input types: (1) metadata about the available windows, (2) a semantic scene description including identified flat surfaces, and (3) user behaviour data such as verbal requests, head direction, and pointing gestures. Without any fine-tuning, the LLM is prompted via a structured system description (shown in Figure~\ref{lst:llm-input}) to act as an assistant that helps position windows on the detected surfaces. The LLM generates a JSON output consisting of actionable placement commands (e.g., \texttt{place}, \texttt{remove}) that are interpreted and executed by the system as simple commands. A visual overview of the architecture is shown in Figure~\ref{fig:system-architecture}, highlighting the integration of scene understanding, user behaviour, and the window workspace with the LLM. %\textcolor{red}{We implemented the system in Unity @@VERSION@@ and tested on Meta Quest 3.}

\subsection{Scene Semantic Understanding}

\begin{figure}[t]
    \centering
    \includegraphics[width=1\columnwidth]{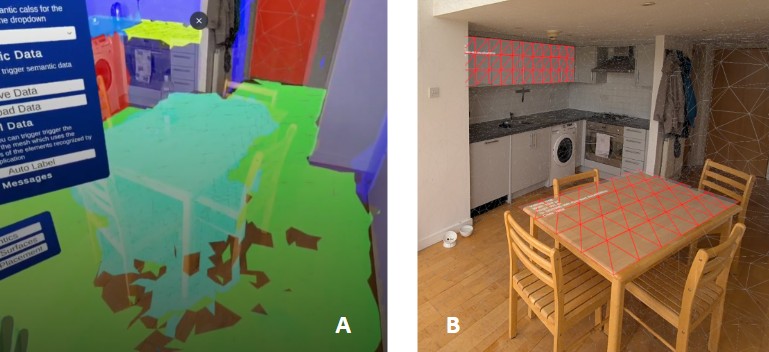}
    \caption{(A) Semantic segmentation of the scene using the Meta Quest API, showing identified classes such as floor, cabinet, and table. (B) Flat surface detection for placing virtual windows, with mesh overlays illustrating usable planar regions.}
    \label{fig:components}
\end{figure}

A crucial component of our system is semantic scene understanding, which allows virtual windows to be contextually placed on appropriate flat surfaces in the user's physical environment. While recent AR research has introduced several advanced methods for semantic segmentation~\cite{rozenberszki2022language,adamyan2023smaller3d}, our system uses a hybrid approach that combines automatic and manual segmentation. Specifically, we used the Meta Quest Scene API~\cite{metaquestprosceneapi2022}, which is natively supported in Unity and available on Quest headsets. This API provides automatic mesh labeling for a limited set of semantic classes (approximately 10), including \textit{wall}, \textit{floor}, \textit{cabinet}, and \textit{table}. To address the limited semantic granularity of automatic labels, we augment the system with manual segmentation capabilities, allowing the user to annotate the scene with additional 30 semantic classes. Table~\ref{tab:semantic-classes} shows the default class set provided by the Meta Quest Scene API, as well as illustrates the extended list and the process of manually re-inspection of the segmentation, such as labeling specific parts of the kitchen floor Figure~\ref{fig:components} (A). This richer semantic understanding enables for more precise and context-sensitive placement of virtual windows by the LLM.

\begin{table}[h]
\centering
\scriptsize
\caption{Semantic segmentation classes used in the system.}
\label{tab:semantic-classes}
\begin{tabular}{|p{0.20\textwidth}|p{0.20\textwidth}|}
\hline
\textbf{Meta Quest API (Automatic)} & \textbf{Extended Manual Labels} \\ \hline
Wall, Floor, Cabinet, Bed, Chair, Sofa, Table, Door, Window, Bookshelf
& 
Picture, Counter, Blinds, Desk, Shelves, Curtain, Dresser, Pillow, Mirror, Floor Mat, Clothes, Ceiling, Books, Refrigerator, Television, Paper, Towel, Shower, Box, Whiteboard, Person, Nightstand, Toilet, Sink, Lamp, Bathtub, Bag, Other Structure, Other Furniture, Other Prop \\ 
\hline
\end{tabular}
\end{table}

\begin{figure*}[t]
  \centering
  \includegraphics[width=0.9\linewidth]{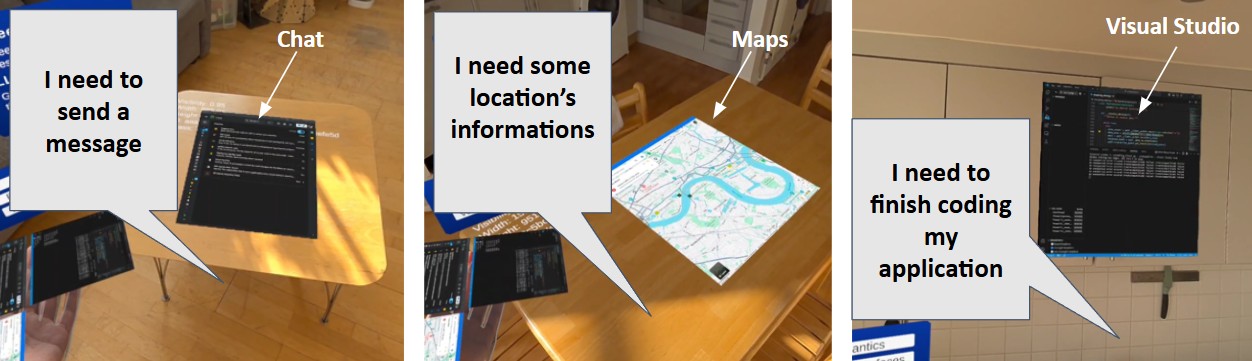}
  \caption{Task–centric window placement.  
  Instead of naming specific applications or surfaces, users simply state their goals (e.g., \emph{“I need to send a message,” “I need some location’s information,” “I need to finish coding my application”}).  
  The LLM interprets each high-level task and (1) selects the relevant window, \emph{Chat}, \emph{Google Maps}, or \emph{Visual Studio}, and (2) places it on an appropriate, visible surface.  
  This allows users to think in terms of \emph{what} they want to accomplish rather than \emph{how} to manage windows, streamlining workflow in XR.}
  \label{fig:task-centric}
\end{figure*}

\subsection{Flat Surface Identification}

Anchoring virtual screens to flat surfaces in real world significantly improves spatial consistency, usability, and safety in immersive XR environments. Previous research shows that placing XR content on familiar, stable surfaces reduces cognitive load, supports intuitive interactions, and mitigates motion sickness and eye strain~\cite{davari2024towards,stanney2021extended,natarajan2024ergonomic,zhang2024towards}. By aligning virtual windows with predictable physical surfaces, users benefit from reduced spatial disorientation and improved ergonomic comfort. To identify suitable planar regions, our system leverages previously computed semantic segmentation and 3D geometric analysis. As illustrated in Figure~\ref{fig:components} (B), the algorithm groups adjacent mesh faces based on coplanarity, evaluated by the dot product of face normals against a fixed angular threshold. Principal Component Analysis (PCA) further refines these planar areas by estimating their dominant orientations, ensuring accurate alignment of virtual content with physical geometry. Identified planar regions inherit semantic labels from the underlying segmentation (e.g., \textit{table}, \textit{cabinet}), allowing semantically informed layout decisions. Each region is represented as a JSON structure, forming the spatial context provided to the LLM for reasoning about the appropriate placement of the windows (Listing~\ref{lst:llm-input}).

\subsection{Window Workspace}
We build on top an existing open-source multi-window XR environment \emph{WindowMirror} that captures existing windows from a PC and renders them inside the XR environment \cite{bovo2024windowmirror}. In addition to the scene description, the LLM also receives a list of available virtual windows. Each window is described as a JSON object with its \texttt{id}, \texttt{size}, \texttt{location}, and \texttt{name}. The \texttt{location} field indicates the surface where the window is currently placed or is set to \texttt{"none"} if the window is not currently visible in the scene. An example of this structure is shown in Listing~\ref{lst:llm-input} \texttt{"windows"}. Each of these windows can be referenced by the LLM when generating placement actions. 

\subsection{Multimodal Input Interpretation}
A key feature of our system is its ability to interpret multimodal user input, combining verbal requests with behavioural cues such as pointing gestures and head direction. Users may issue underspecified commands like ``Can you move this here?’’, a common form of deictic or pointing communication. To disambiguate such commands, the system supplements speech input with a list of recent pointing events, which includes identifiers of hovered objects and the duration of the hover. This information helps the LLM infer the user's intended target window and surface. Additionally, head direction is used to compute the visibility of each surface at the moment of the verbal command. As shown in Listing~\ref{lst:llm-input}, the LLM receives input with fields like \texttt{"flat\_surface"} and \texttt{"visibility"}, this visibility score is used by the LLM to prioritize where windows should be placed, favouring those surfaces that are most visible to the user at the time of the request. This enables a richer interpretation of ambiguous language through alignment with implicit behavioural signals.

\lstdefinelanguage{json}{
    basicstyle=\scriptsize\ttfamily,
    numbers=left,
    numberstyle=\tiny,
    stepnumber=1,
    numbersep=5pt,
    showstringspaces=false,
    breaklines=true,
    frame=single,
    backgroundcolor=\color{gray!10},
    literate=
     *{0}{{{\color{blue}0}}}{1}
      {1}{{{\color{blue}1}}}{1}
      {2}{{{\color{blue}2}}}{1}
      {3}{{{\color{blue}3}}}{1}
      {4}{{{\color{blue}4}}}{1}
      {5}{{{\color{blue}5}}}{1}
      {6}{{{\color{blue}6}}}{1}
      {7}{{{\color{blue}7}}}{1}
      {8}{{{\color{blue}8}}}{1}
      {9}{{{\color{blue}9}}}{1},
}

\begin{lstlisting}[language=json,caption={Prompt design.},label={lst:prompt-design}]]
  "description": "You are an assistant that answers in JSON format {\"response\":\"your-response\",\"actions\":\"supported-actions\"}. Your purpose is to help me organize my virtual windows on the available flat surfaces around my office/house/space. You will do so by generating actions that are then executed by the system.",
  "action_format": {
    "actions": [
      ["place", "windowid", "surface"],
      ["remove", "windowid", "surface"]
    ]
  },
  "inputs": {...},
  "task": "Interpret the user's request to place or remove one or more windows on the appropriate surfaces to maximize visibility, match the user request, and consider inferred preferences. Note that windows are automatically resized, but depending on the flat surface size, you might not want to cramp too many or too few of them.",
  "example_1": {
    "input": {...},
    "expected_output": {...}
  },
  "pointing_behavior_usage": "Use pointing behavior as a nonverbal cue to integrate the user request. If language is ambiguous, use pointing to clarify and resolve references.",
  "example_2": {
    "input": {...},
    "expected_output": {...}
  },
  "notes": "Ignore pointing events with very short hover durations as they may be noise from the user passing over objects. Exclude the current surface of a window from target selection when interpreting pointing behavior."
\end{lstlisting}

\noindent\textbf{Prompt design strategies:}  
Prompts are structured to clearly define the LLM’s role, output format, and decision-making constraints. The system message specifies that the assistant must always respond in JSON with a \texttt{"response"} field and an \texttt{"actions"} array of \texttt{[action, windowid, surface]} triplets. The prompt includes a detailed description of the task, the semantic meaning of each input field (\texttt{user\_request}, \texttt{flat\_surfaces}, \texttt{windows}, \texttt{userPointingEvents}), and explicit examples of input–output pairs for both unambiguous and ambiguous scenarios. To improve grounding and reduce hallucinations, the prompt also encodes rules for resolving deictic references using pointing behavior, prioritizing visible surfaces, avoiding redundant placements, and filtering out short hover events as noise. This combination of role specification, structured schema, behavioral constraints, and worked examples guides the LLM toward consistent, context-aware action generation. An excerpt of the full system prompt is shown in Listing~\ref{lst:prompt-design}. All prompt processing and JSON action generation were performed using OpenAI’s GPT-4 (March 2025) via the Chat Completions API.

\section{Discussion}

The original "Put-That-There" was a landmark in direct manipulation, proving that a computer could understand ambiguous, multimodal context-dependent commands \cite{10.1145/800250.807503}. Building upon, yet moving beyond, this legacy of direct manipulation seen in subsequent window placement research \cite{pavanatto2025spatial, satriadi2020maps, lee2020shared}, our work highlights the transformative potential of integrating LLMs into multimodal XR window management. By enabling an AI agent to handle complex reasoning, we shift interaction from explicit commands to higher-level abstractions, where the user simply specifies a goal. In the following subsections, we discuss the implications of moving from direct manipulation commands to goal oriented commands which can generate a series of window(s) placement action(s). Together, these discussions underscore opportunities and open challenges in leveraging AI-driven multimodal interactions for XR productivity.

\begin{lstlisting}[language=json,caption={Example of LLM input JSON structure.},label={lst:llm-input}]
{"userPointingEvents": [
    {
      "identifier": "e5f3b127...",
      "hoverDuration": 1.5
    },
    { ... }
  ],
  "windows": [
    {
      "id": "e5f3b127...",
      "size": "200x200",
      "location": "none",
      "name": "Google Maps"
    },
    { ... }
  ],
  "flat_surfaces": [
    {
      "id": "7409038c...",
      "size": "500x700",
      "visibility": 0.8,
      "semantic": "cabinet",
      "current_windows": []
    },
    { ... }
  ]}
\end{lstlisting}

%\begin{lstlisting}[language=json,caption={Example LLM output JSON for an explicit request.},label={lst:llm-output}]
%{
%  "response": "Actions generated to place Google Maps on the Table.",
%  "actions": [
 %   ["place", "Google Maps", "Table"]
%  ]
%}
%\end{lstlisting}

\begin{lstlisting}[language=json,caption={Example LLM output JSON for a goal-centric request triggering multiple actions.},label={lst:llm-output-multiple}]
{"response": "Actions generated to place Google Maps, Notes, and Calendar on visible surfaces for trip planning.",
  "actions": [
    ["place", "Google Maps", "Table"],
    ["place", "Notes", "Desk"],
    ["place", "Calendar", "Wall"]
  ]}
\end{lstlisting}

\subsection{From One-to-One to One-to-Many}

Explicit window placement modalities typically follow a one-to-one mapping, each user action results in a single, specific placement ~\cite{pavanatto2025spatial}. In contrast, LLM-based interactions support a one-to-many model, where a single high-level request can trigger multiple coordinated actions, such as selecting relevant windows, repositioning them, or reorganizing the entire workspace. This shift enables more efficient, goal-driven workflows but introduces new questions around user control, transparency, and alignment between user intent and system behaviour. While this automation reduces manual effort, it may also require new forms of feedback and affordances to maintain user trust and understanding.

% add example here 

\subsection{Goal-Centric Window Placement}

Traditional XR interfaces require users to explicitly specify both the application and its placement, actions that are often cumbersome and cognitively demanding in immersive environments. In contrast,LLMs empower a higher level of abstraction: users can simply express their goals (e.g., \emph{“I need to send a message”}, \emph{“I need some location’s information”}, or \emph{“I need to finish coding my application”}) without naming specific applications or targets. The LLM interprets these high-level intents and maps them to the appropriate applications, such as Chat, Google Maps, or Visual Studio, based on semantic reasoning over the task description and available windows.

This type of flexible goal-to-application mapping is made possible through the generative and contextual reasoning capabilities of LLMs. As shown in Figure~\ref{fig:task-centric}, the system selects the relevant application and places it on a suitable visible surface, streamlining the user’s workflow and shifting interaction from...

\subsection{Application Interrelationships Modelling}

Beyond simple goal-to-application matching, LLMs can reason about the semantic relationships between multiple applications to support more coordinated and context-aware window management. For example, if a user says \emph{“I need to find images for a presentation,”} the system infers the need to open both a browser for research and a slide editor for content creation. This understanding of functional dependencies allows the LLM to launch and arrange multiple relevant windows as a cohesive workspace, rather than treating each application in isolation.
Such relational reasoning is challenging to encode with traditional rule-based systems, but LLMs can dynamically infer task structures and workflows from natural language. This capability supports richer, more adaptive interaction paradigms in XR, where productivity often spans multiple tools and contexts simultaneously.

%% interellation between 

\section{Future Work and Conclusion}

A critical next step involves evaluating the proposed system through user studies to empirically assess its impact on cognitive load, task performance, and user experience. Such evaluations will compare our multimodal, LLM-driven window management against traditional explicit manipulation approaches. This will provide insights into how effectively goal-centric and automated interactions reduce cognitive overhead, influence task efficiency, and affect user trust and satisfaction in immersive productivity scenarios. We introduced a multimodal, LLM-driven system for task-centric window management in XR. By combining verbal input with pointing, head gaze, and scene semantics, the system enables high-level, goal-oriented interactions that reduce the need for manual window manipulation. This shift from one-to-one commands to one-to-many, intent-driven actions supports more efficient and adaptive XR workflows. 

%\subsection{Bibliography Instructions}

%%\acknowledgments{
%The authors wish to thank A, B, and C. This work was supported in part by
%a grant from XYZ.}

%\bibliographystyle{abbrv}
\bibliographystyle{abbrv-doi}

\bibliography{template}

\end{document}